\pdfoutput=1
\documentclass[10pt,letterpaper]{article}
\usepackage{opex3}

\begin{document}

\title{Photonic crystal formation on optical nanofibers using femtosecond laser ablation technique}

\author{K. P. Nayak$^{\ast}$ and K. Hakuta}

\address{Center for Photonic Innovations, University of Electro-Communications,\\ Tokyo 182-8585, Japan}

\email{$^{\ast}$kali@cpi.uec.ac.jp} 



\begin{abstract*}
We demonstrate that thousands of periodic nano-craters are fabricated on a subwavelength-diameter tapered optical fiber, an optical nanofiber, by irradiating with just a single femtosecond laser pulse. A key aspect of the fabrication is that the nanofiber itself acts as a cylindrical lens and focuses the femtosecond laser beam on its shadow surface. We also demonstrate that such periodic structures on the nanofiber, act as a 1-D photonic crystal (PhC). Such PhC structures on the nanofiber will strongly enhance the field confinement in such a tapered fiber-based system and may open new avenues in nanophotonics and quantum information technology.\\
\end{abstract*}



\section{Introduction}

Confinement of light to wavelength-scale volumes possesses important
implications to control the light-matter interactions at single photon and
single emitter level. The emerging field of nanofabrication technology has
encouraged various designs of micro/nano-photonic structures to be
investigated. In particular 2-D \cite{scherer pcm} and 1-D \cite{foresi,
painter nb} photonic crystal (PhC) structures have enabled new prospects in
nanolasers \cite{painter}, chemical sensing \cite{loncar sensing} and
optical switching \cite{tanabe}. Moreover, the strong light-matter coupling 
\cite{nature pcm, imamoglu pcm, dirk, lukin pcm} in these PhC structures,
has opened new avenues for quantum information technology. However,
integrating these PhC structures to fiber-based communication channels
still remains challenging. In this context, tapered optical fibers with subwavelength diamters, known
as optical nanofibers, may provide a promising platform for realizing strong light-matter
interaction in a fiber-based system. 

In recent years, optical nanofibers have
opened a new paradigm in manipulating single/few-atom fluorescence \cite%
{opex, njp, chandra opex}. A nanofiber is realized by adiabatically tapering
a conventional single mode optical fiber which enables light in the
fundamental mode to be coupled in and out of the nanofiber with more than $%
95\%$ efficiency. Due to the subwavelength diameter the field in the guided
mode is strongly confined in the transverse direction. As a result the
spontaneous emission of atoms can be strongly modified around the nanofiber
and an appreciable amount of atomic fluorescence can be coupled to the
guided modes \cite{spont em, chandra prl}. Also the strong evanescent tail
of the guided field gives excellent access for trapping and probing few
atoms around the nanofiber \cite{scattering, trap1, arno, kimble trap}.

The light-matter coupling in such nanofiber system can be substantially
improved by fabricating a 1-D PhC structure on the nanofiber. The PhC
structure provides longitudinal confinement of the field in the nanofiber
guided modes. Due to the strong transverse confinement of the nanofiber
guided modes, the coupling between the atom and the guided modes will be
significantly enhanced and the strong-coupling regime can be realized, even
for moderate finesse \cite{cavity coupling}. Such PhC nanofiber system may
become a promising workbench for quantum non-linear optics and may serve as
nodes in quantum networks \cite{kimble}. Also the PhC nanofibers
may open up exciting new applications in lasing, optical switching and chemical/biological sensing.

In this paper, we present the fabrication of PhC nanofibers using
a femtosecond laser ablation technique. We demonstrate that thousands of periodic
nano-craters are fabricated on optical nanofibers by irradiating with just a
single femtosecond laser pulse. A key aspect of the fabrication is the
lensing effect of the nanofiber itself. We also demonstrate that such
periodic nano-craters on the nanofiber, induce strong modulation of refractive index 
and act as a 1-D PhC.

\section{Experiments}

\subsection{Fabrication setup}

\begin{figure}[tbph]
\begin{center}
\includegraphics[height=8.5cm, width=8.5cm]{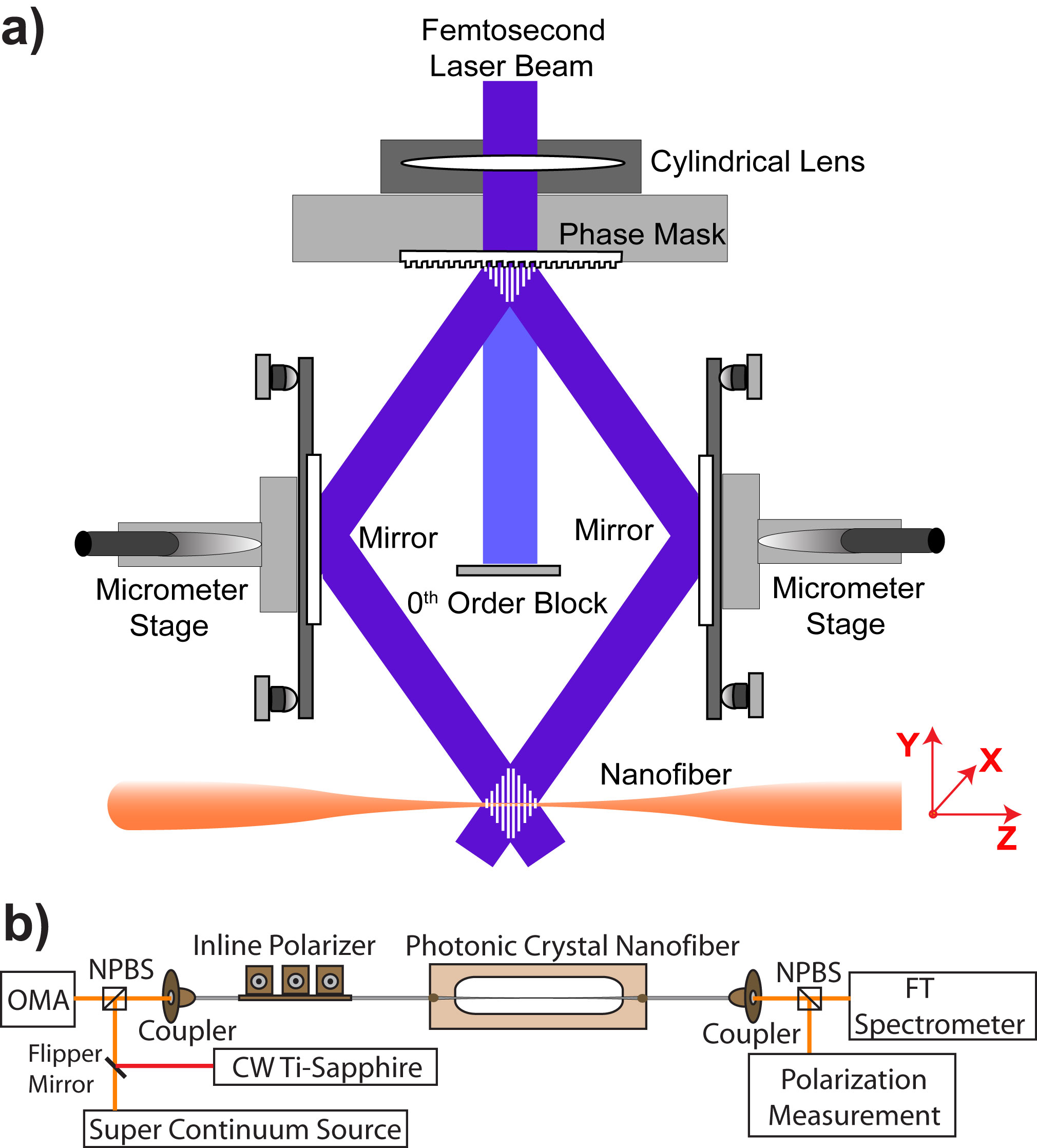}
\end{center}
\caption{(Color online) (a) The schematic diagram of the fabrication setup. The phase
mask splits the femtosecond laser beam into $\pm $ first orders which are
then recombined by the folding mirrors to create an interference pattern at
the nanofiber. A
cylindrical lens is used to line focus the femtosecond laser along the
nanofiber. A zero order block is used to avoid any residual zero order light
in the interference region. (b) The schematic diagram of the optical measurements. 
The transmission and reflection spectra of the fabricated nanofiber samples are simultaneously measured by varying the 
polarization of the input light (see text for details). NPBS 
and OMA denote non-polarizing beam splitter and optical multi-channel analyzer, respectively.}
\label{fig1}
\end{figure}

A schematic diagram of the fabrication setup is shown in Fig. 1(a). The
femtosecond laser (Coherent Libra-HE) used for the fabrication has a central
wavelength ($\lambda $) of 400 nm and generates 120 fs pulses at a
repetition rate of 1 kHz with maximum pulse energy of 1.3 mJ. A Talbot
interferometer \cite{DUV}, consisting of a phase mask as beam splitter and
two folding mirrors, is built to create the two-beam interference pattern on
the nanofiber. The phase mask used for the fabrication has a uniform pitch ($%
\Lambda _{P}$) of 700 nm and was designed for zero order supression at 400
nm wavelength. The position of the folding mirrors is carefully chosen to
symmetrically recombine the first orders at the nanofiber position, thereby
creating an interference pattern with a period of $\Lambda _{P}/2=350$ nm. 
A cylindrical lens is used to line-focus the femtosecond
laser beam along the nanofiber. The typical 1/e$^{2}$ beam size at the
nanofiber is 5.4 mm $\times $ 60 $\mu $m. The polarization of the
femtosecond laser beam is perpendicular to the nanofiber axis. Tapered
nanofibers with waist diameters ($2a$) of 450-650 nm are used for the
fabrication. After
the fabrication, the nanofiber samples are observed using a scanning electron
microscope (SEM).

\subsection{Alignment of the fabrication setup} 
The interferometer alignment is highly sensitive to the optical path length
difference bewteen the two first orders. The path length difference must be
smaller than 36 $\mu $m as the spatial overlap is limited by the femtosecond
laser pulse width \cite{DUV, femtoreview1}. Hence, the positions of the
folding mirrors are controlled by micrometer stages and the mirror tilt
angles are precisely controlled. The alignment of the interferometer is
performed by observing the periodic ablation pattern on a glass plate, using
the SEM. After alignment of the interferometer, the
glass plate is replaced by the tapered nanofiber. For the fabrication, the
tapered fiber is mounted on a metallic holder having a central hole of 6 cm
length, so that the nanofiber is protected from contamination due to
ablation of the substrate. The overlap of the femtosecond laser beam with
the nanofiber is monitored by reducing the pulse energy to minimum value and
observing the scattered light from the nanofiber using a CCD camera. Maximum
overlap is achieved using a X-stage and a rotation-stage.

\subsection{Measurement of optical properties} 
The optical properties of the
fabricated samples are characterized by measuring the transmission and
reflection spectra. A schematic of the
experimental setup is shown in Fig. 1(b). A supercontinuum light source (SuperK Extreme, NKT
Photonics) is launched into the tapered fiber sample and the spectrum of the
transmitted light is measured using a Fourier transform spectrometer
(FT-spectrometer). The resolution of the FT-spectrometer (Nicolet 8700,
Thermo Fisher Scientific) is $\sim 0.125$ cm$^{-1}$($\sim 0.01$ nm).
An inline fiber polarizer is used to control the polarization of the input
light. For confirmation, a part of the transmitted light is used for
polarization measurements. Simultaneously, the spectrum of the reflected
light is measured using an optical multi-channel analyzer (OMA). The
resolution of the OMA (QE65000, Ocean Optics) is $\sim 2$ nm. The
transmittance and reflectance values are calibrated using a CW Ti-Sapphire
laser source (MBR-110, Coherent Inc.).

\section{Results and Discussions}

\subsection{Nano-crater formation on nanofibers}

For the fabrication, we vary the pulse energy, repitition rate and
irradiation time (number of shots) to find the optimum conditions. It was found that for 1 kHz repitition rate, the
nanofiber can be completely destroyed within 1 second, even at pulse energy
of 200 $\mu $J. Even a repition rate of 100 Hz can still seriously damage the nanofiber. 
However, we have found that by reducing the number shots to less than 20-shots, one could realize
controlled nanofabrication on the nanofiber. By systematically changing the number
of shots and pulse energy, we have found that single-shot irradiation is the
best condition to realize clean nanofabrication. In the following, we discuss the fabrication results
for multiple-shot and single-shot irradiation. 

\subsubsection{Multiple-shot fabrication}

Figure 2(a) shows the SEM image of a
typical sample fabricated by 20-shot irradiation with a pulse energy of 370 $%
\mu $J. We have found that many ablated structures are formed on the nanofiber. 
We must mention that these ablated structures are
formed, not on the irradiation side, but on the shadow side of the
nanofiber. As one can see the ablation pattern is quite irregular. However, a crucial observation is
that although the beam size along X-axis (60 $\mu $m) is much larger than
the nanofiber diameter, the ablation pattern is formed in a line along the
fiber axis (Z-axis). 

Figure 2(b) shows the SEM image of a
typical sample fabricated by 3-shot irradiation with a pulse energy of 560 $%
\mu $J. One can see rather regular and periodic ablation pattern is formed on 
the shadow side of the nanofiber.
However, the ablated structures are elliptical and are elongated along the Z-axis.
The typical size of an ablated structure is 90 nm $\times $ 210 nm. We must mention that
such ellipticity is observed throughout the ablation pattern, even for the weakest ablated structure.

\begin{figure}[tbph]
\begin{center}
\includegraphics[width=12.5cm]{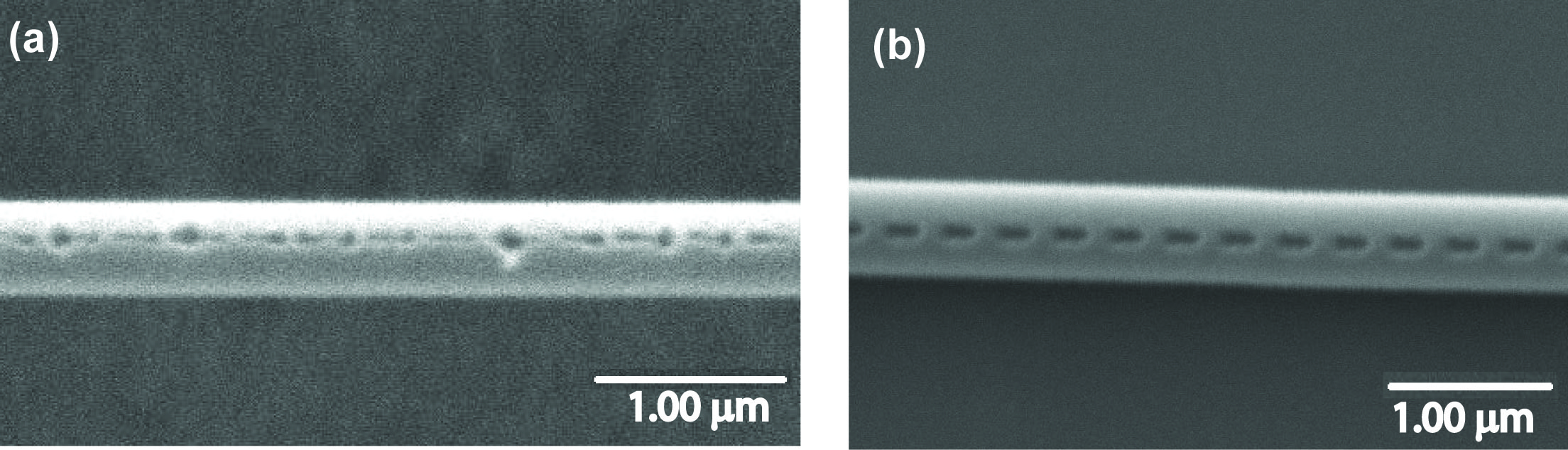}
\end{center}
\caption{(Color online) SEM images of samples fabricated by multiple-shot irradiation. 
(a) SEM image of a typical
sample fabricated by 20-shot irradiation. (b) SEM image of a typical
sample fabricated by 3-shot irradiation. 
The ablated structures are observed on the shadow side of the nanofiber.}
\label{fig2}
\end{figure}

\subsubsection{Single-shot fabrication}

\begin{figure}[tbph]
\begin{center}
\includegraphics[width=12.5cm]{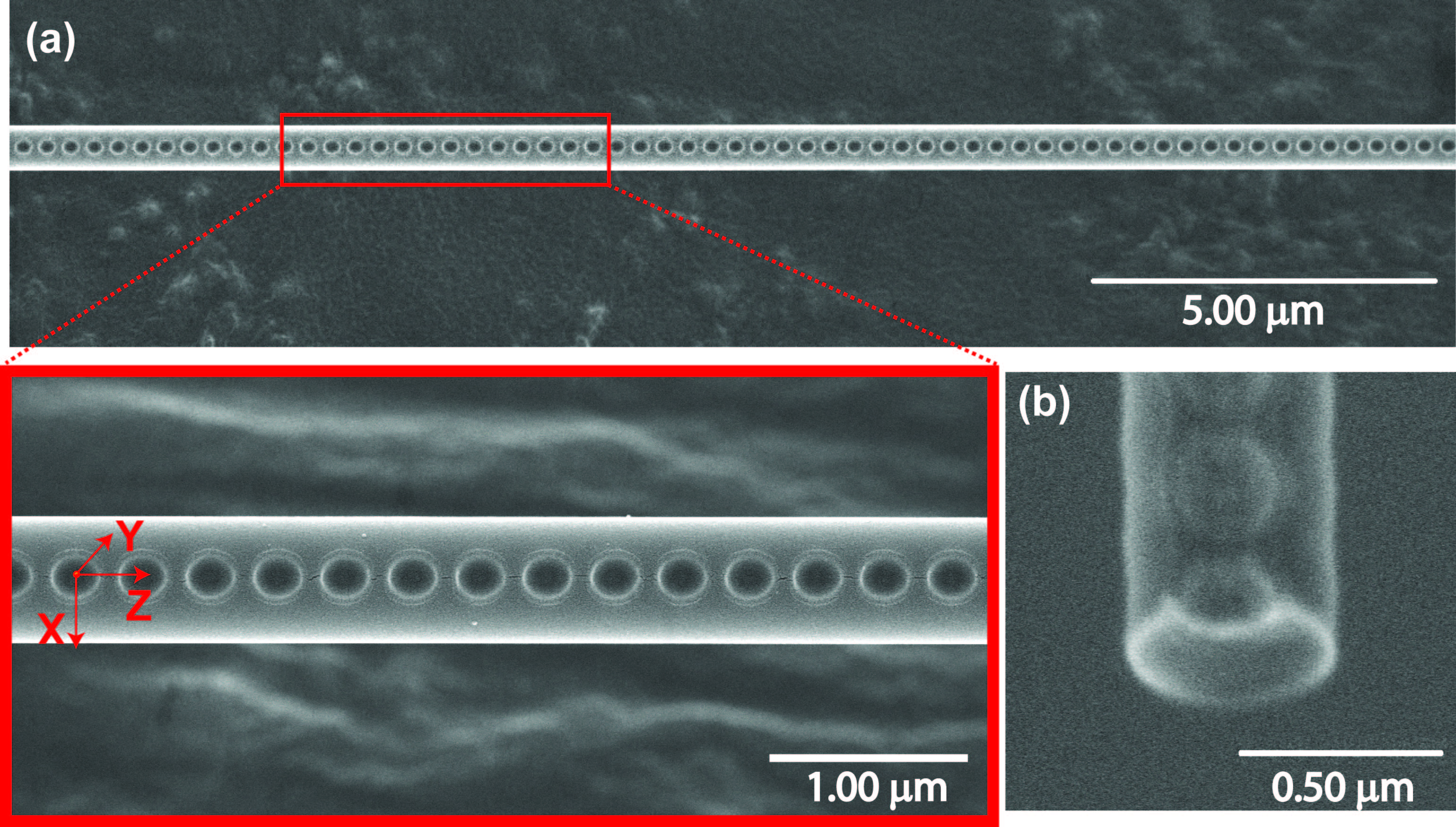}
\end{center}
\caption{(Color online) (a) SEM image of a typical sample
fabricated using single-shot irradiation. The inset shows the enlarged view. 
The periodic nano-crater
structures are observed on the shadow side of the nanofiber. (b)
The cross-sectional image of a typical nano-crater measured by tilting the
nanofiber at an angle of 33$^{\circ }$.}
\label{fig2}
\end{figure}

Figure 3(a) shows the SEM image of a typical sample fabricated using
single-shot irradiation with a pulse energy of 630 $\mu $J. As one can
clearly see, periodic nano-crater structures are formed on the shadow side of the nanofiber.
Such periodic structures are quite systematically formed over a length of
few mm, consisting of thousands of such nano-craters. The inset shows the enlarged 
view of the sample. The shape of the
nano-craters is almost circular and the diameter of a typical nano-crater is
around 210 nm. The nano-craters are formed with a periodicity of 350 nm,
which corresponds well to the interference fringe spacing. The depth of a
typical nano-crater is measured by cutting the nanofiber at the nano-crater
position and observing the cross-section using the SEM. The cross-sectional
view is shown in Fig. 3(b). The nano-crater has a bowl-like shape and the
depth is $\sim 120$ nm.

\subsubsection{Lensing effect of the nanofiber}

It should be noted that the ablated structures shown in Fig. 2 and Fig. 3 are
formed on the shadow side of the
nanofiber. We could not observe any particular structure on the irradiation
side. This suggests that the nanofiber itself acts as a cylindrical
lens and focuses the femtosecond laser beam on its shadow surface \cite%
{book1,trap}. Also as shown in Figs. 2(a) and (b), the ablation pattern is formed
exactly in a line along the fiber axis (Z-axis), confirming the lensing
effect of the nanofiber. The uniqueness of the present method is that the
lensing effect of the nanofiber makes it robust to any mechanical
instabilities in the transverse direction (X-axis). But 
the instabilities along Z-axis can seriously affect the fabrication for multiple-shot irradiation, as the intensity pattern on the 
nanofiber may differ for each shot. This is evident from the observed irregular ablation pattern for
20-shot irradiation (Fig. 2(a)), where the periodicity is completely washed out due to the instabilities along Z-axis. 
Even for 3-shot irradiation (Fig. 2(b)) one can see that the ablated structures are elongated along the 
fiber axis. However such instability does not affect the
fabrication in single-shot irradiation as the irradiation time is only 120
fs (i.e. pulse length). As a result periodic nanostructures with well
defined shape (shown in Fig. 3(a)) and periodicity are fabricated, without
taking any special care to suppress mechanical vibrations. Moreover, such single-shot irradiation 
technique makes the fabrication method highly reproducible.
The lensing effect of the nanofiber may
explain the shape of the nano-craters. Although the field distribution along
the Z-axis is limited by the interference pattern, the circular shape of the
nano-craters is mainly due to focusing of the femtosecond laser along the
X-axis.

\subsection{Diameter distribution of nano-craters fabricated using single-shot irradiation}

\begin{figure}[tbph]
\begin{center}
\includegraphics[width=7.5cm]{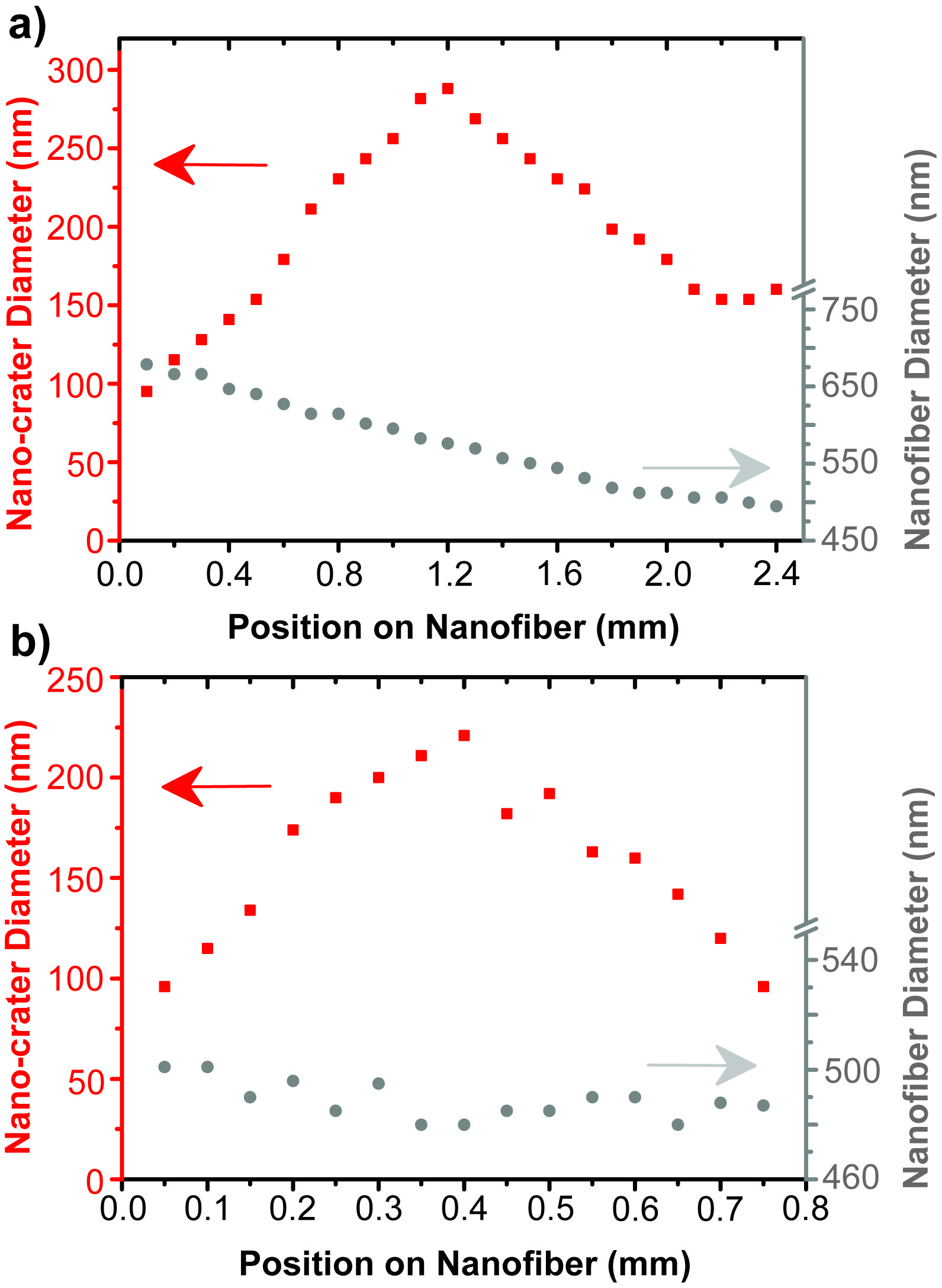}
\end{center}
\caption{(Color online) The diameter distribution of the nano-craters for two nanofiber samples, (a) Sample 1 and (b) Sample 2, 
fabricated using single-shot irradiation. The red squares denote the diameter of nano-craters
and the gray circles denote the corresponding diameter of the nanofiber.}
\label{fig3}
\end{figure}

The diameter distribution of the nano-craters for two nanofiber samples, 
fabricated using single-shot irradiation, are
plotted in Figs. 4(a) and (b), along with the corresponding nanofiber diameter. 
For Sample 1 the nano-craters are fabricated in the
tapered region, and the diameter of the nanofiber varies from 500 nm to 680
nm. A periodic array of nano-craters is formed in a region exceeding 2 mm
along the nanofiber axis. The diameter profile of the nano-craters shows a
peak-like behavior and the 1/e$^{2}$ width is $\sim 2.6$ mm. The diameter of
the nano-craters varies from 95 nm to 290 nm. On the other hand, for Sample
2 the fabrication is done near to the waist region, and the diameter of the
nanofiber is almost constant around 490 nm. The nano-craters are formed in a
region of 0.8 mm along the nanofiber. The diameter of the nano-craters
varies from 95 nm to 220 nm and the 1/e$^{2}$ width of the profile is $\sim
1.03$ mm. The observations also suggest rather smaller depth of nano-craters
for Sample 2 as compared to the Sample 1.

The observed 1/e$^{2}$ width of the diameter profile of the nano-craters is
smaller than the original beam profile. This is due to the fact that the
femtosecond laser induced ablation involves a multi-photon ionization
process \cite{femtoreview1, type2}. Hence the ablation process scales as a
higher power of the intensity and has some threshold condition. In order to understand the diameter 
profile of the nano-craters one must also consider the nanofiber diameter, as the lensing effect of the 
nanofiber depends on the size parameter ($2\pi a/\lambda $). In case of
Sample 2, the reduction in the length of the nano-crater array and the
smaller depths of the nano-craters may be understood from the decrease in
the lensing effect of the nanofiber due to thinner diameter \cite{book1,trap}. The observation of nano-crater diameters down to 95 nm suggests that
one can design various nanofabrications. The diameter distribution of the nano-craters
can be designed by controlling the intensity distribution and the size
parameter.

\subsection{Optical properties of nanofibers with periodic nano-craters}

\begin{figure}[tbph]
\begin{center}
\includegraphics[width=8.0cm]{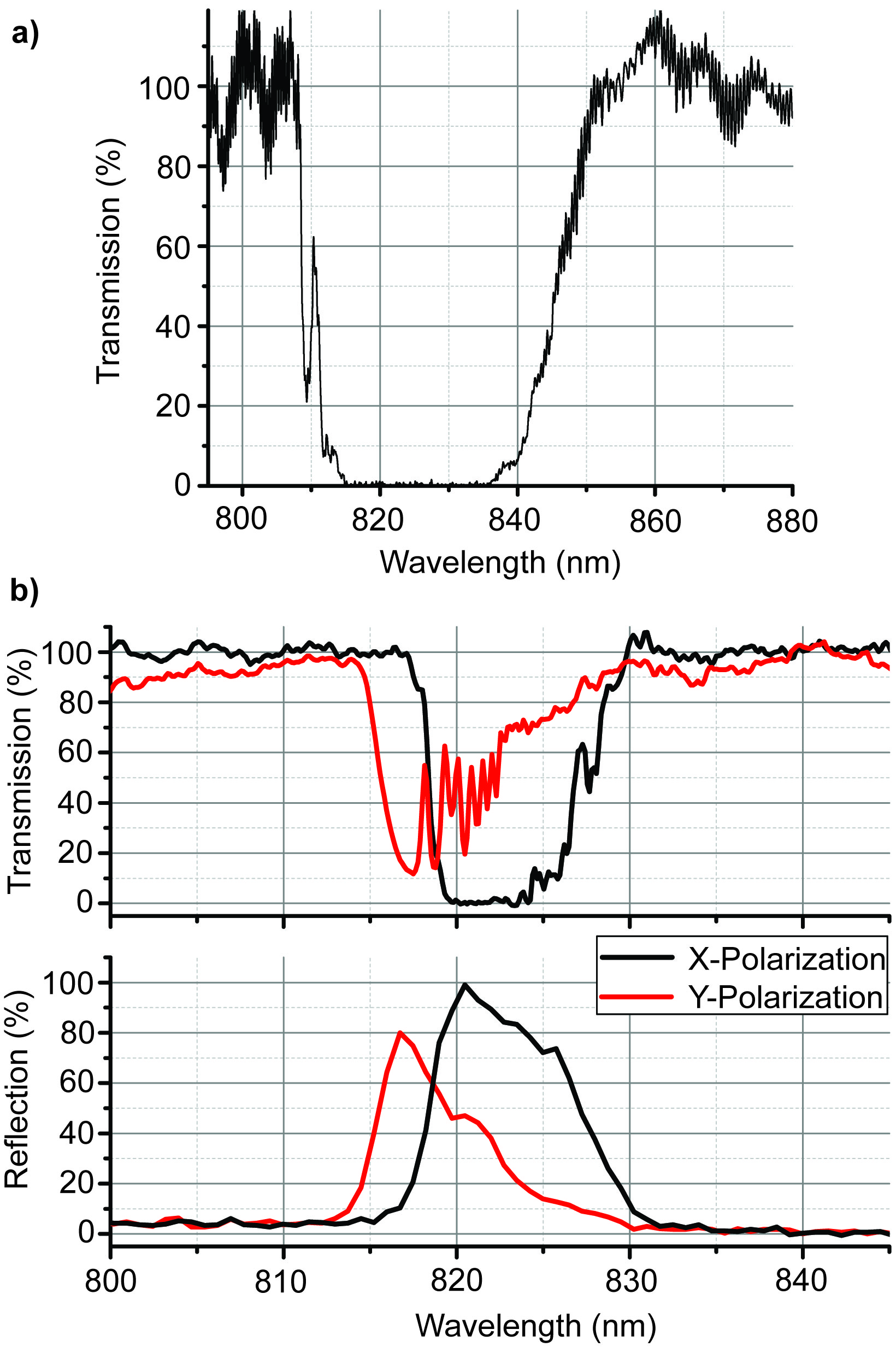}
\end{center}
\caption{(Color online) Optical characteristics of nanofiber samples fabricated using 
single-shot irradiation. (a) The transmission
spectrum of Sample 1. (b) The transmission and reflection spectra of Sample
2, measured for two orthogonal polarizations, X-polarization (black curves)
and Y-polarization (red curves).}
\label{fig4}
\end{figure}

Optical properties of the nanofiber samples, fabricated using 
single-shot irradiation, are characterized by measuring the transmission and reflection spectra.
The transmission spectrum of Sample 1, shown in Fig. 5(a), shows a strong,
broad reflection band centered around 825 nm. The FWHM of the reflection
band is around 37 nm. In the wavelength region from 815 nm to 836 nm there
is almost no transmission, indicating the stop band. The reflectivity values
measured using a CW Ti-Sapphire laser confirms high reflectivity of $99\%$
in this stop band region. The transmission and reflection spectra of Sample
2 are shown in Fig. 5(b). The transmission and reflection spectra are
measured for two orthogonal linear polarizations, X-polarization (along
X-axis) and Y-polarization (along Y-axis). The transmission spectrum for the
X-polarization shows a reflection band centered around 823 nm and the FWHM
is 10 nm. The wavelength region from 819.5 nm to 824 nm denotes the stop
band. Note that the transmission spectrum for the Y-polarization is blue
shifted having a reflection peak at 817.5 nm and FWHM of 6.8 nm. The
spectrum is asymmetric and shows several small sharp peaks in the red side.
The reflection spectra almost match to the reflection band in the
transmission spectra except that the fine features are washed out due to the
resolution limit for the reflection measurements. The peak reflectivity
values are measured using a CW Ti-Sapphire laser. For the X-polarization, we
could observe almost 99\% reflection in the stop band region, whereas for the
Y-polarization the peak reflectivity is measured to be 80\%.

The observation of high reflectivity values in the stop band region, clearly demonstrates that the periodic nano-craters on the
nanofiber induce strong modulation of refractive index and act as a 1-D PhC.
The observed spectral width for Sample 1 is $3.7$ times broader than the
Sample 2. This is mainly due to the nanofiber diameter variation for Sample
1, as the diameter determines the propagation constant and the Bragg
resonance \cite{FIB 1, FIB 2}. In case of Sample 2, the observed blue shift
for the Y-polarization may be explained using the same argument \cite{FIB 1,
FIB 2}. Due to formation of nano-craters, the effective diameter along the Y-axis is reduced which results in such blue shift. 
Also the transmission spectrum for the Y-polarization is asymmetric
and shows several small sharp peaks in the red side. Further investigations
are required to understand such fine spectral features. 

The peak reflectivity value for the Y-polarization is measured to be $80\%$ whereas
the transmittance value at this wavelength is $12.5\%$. This suggests that,
for the Y-polarization, there is few percent scattering loss due to the
nano-craters \cite{FIB 1}. However, for the X-polarization, the scattering loss is negligible as is evident 
from the high transmittance values away from the stop-band region. One should expect high
scattering loss as thousands of nano-craters are formed on the nanofiber.
The observed low scattering loss and high reflectivity is mainly due to the shape and shallow
depths of the nano-craters (shown in Fig. 3(b)). The observation of such excellent optical 
properties, clearly suggests that one can design various types of 1-D PhC cavities on nanofibers using the present fabrication method.

\section{Conclusion}

In conclusion we have demonstrated the formation of periodic nano-craters on
a sub-wavelength diameter silica fiber using a femtosecond laser ablation
technique. The ablation is achieved by irradiating only a single femtosecond
pulse. Thanks to the lensing effect of the nanofiber, circular nano-crater
structures are formed on the shadow surface of the nanofiber. Such a
fabrication method may open new prospects in nanophotonics and
nanofabrication technologies. We have demonstrated that the periodic
nano-craters on the nanofiber act as a 1-D PhC. The PhC nanofiber\ system
can become a promising workbench for quantum non-linear optics and will open
new avenues in quantum information technology, by combining with
laser-cooled atoms or solid-state quantum emitters. Also the PhC nanofibers
may open up exciting new applications in lasing, optical switching and chemical/biological sensing.

\section*{Acknowledgements}
We are thankful to Mark Sadgrove and Makoto Morinaga for fruitful
discussions. Also we wish to thank Hitachi High-Tech Corp., Japan, for
helping in SEM measurements with SU8040. This work was supported by the
Japan Science and Technology Agency (JST) as one of the Strategic Innovation
projects.

\end{document}